# Automatic Bug Triage using Semi-Supervised Text Classification

Jifeng Xuan[1]     He Jiang[2,3]     Zhilei Ren[1]     Jun Yan[4]     Zhongxuan Luo[1,2]

[1]School of Mathematical Sciences, Dalian University of Technology, Dalian, 116024 China
[2]School of Software, Dalian University of Technology, Dalian, 116621 China
[3]State Key Laboratory of Computer Science, Institute of Software, Chinese Academy of Sciences, Beijing, 100190 China
[4]Technology Center of Software Engineering, Institute of Software, Chinese Academy of Sciences, Beijing, 100190 China
[1]{xuan, ren}@mail.dlut.edu.cn     [2]{jianghe, zxluo}@dlut.edu.cn     [4]junyan@acm.org

*Abstract*—In this paper, we propose a semi-supervised text classification approach for bug triage to avoid the deficiency of labeled bug reports in existing supervised approaches. This new approach combines naive Bayes classifier and expectation-maximization to take advantage of both labeled and unlabeled bug reports. This approach trains a classifier with a fraction of labeled bug reports. Then the approach iteratively labels numerous unlabeled bug reports and trains a new classifier with labels of all the bug reports. We also employ a weighted recommendation list to boost the performance by imposing the weights of multiple developers in training the classifier. Experimental results on bug reports of Eclipse show that our new approach outperforms existing supervised approaches in terms of classification accuracy.

## I. Introduction

Most of large software projects employ a bug tracking system (bug repository) to manage bugs and developers. In software development and maintenance, a bug repository is a significant software repository for storing the bugs submitted by *users*. Those users, including developers, testers and end users, submit the content of bugs as *bug reports* to identify software defects or user suggestions. In the bug repositories, Bugzilla [6] is the most popular one in open source softwares.

Before verifying and modifying a bug, each bug report must be assigned to a relevant developer who could fix it [7]. This process of assignment is called *bug triage*. In traditional bug repositories, all the bugs are manually triaged by some specialized developers (triagers) [7]. Such manual work is expensive in labor costs. Taken Eclipse (an open source integrated development environment [8]) as an example, 37 bugs are submitted to the bug repository and 3 person-hours are spent on assigning the bug reports per day on average [1].

Aiming to reduce the human labor costs, some supervised text classification approaches have been proposed for automatic bug triage, including naive Bayes (NB) classifier [7] and support vector machine [3]. These approaches treat bug triage as the classification of text content of bug reports and treat relevant developers as class labels. The supervised approaches train learnable classifiers with existing bug reports and then predict relevant developers for the incoming bug reports with these classifiers. In addition, the classification accuracy of these approaches is not high enough, so most of the bug triage approaches employ a recommendation list to provide the candidate developers to be selected by human triagers.

Before training a supervised classifier for bug triage, a necessary step is to collect numerous *labeled bug reports*, which are bug reports marked with their relevant developers. However, labeled bug reports are insufficient. Bettenburg, et al. investigate the existing quality problems of the bug reports in actual bug repositories [4]. Since some of bug reports are not well-formed, it is difficult to provide correct labels for all the bug reports. In practice, even a human triager usually mistakenly labels bug reports with developers who cannot fix the bugs. Jeong, et al. report that 44% of bugs are assigned to mistaken developers by triagers during the first assignment [11]. In other words, nearly half of bug reports may consist of wrong labels after they are assigned. Motivated by the deficiency of labeled bug reports with good quality, we take advantage of *unlabeled bug reports*, which are original bug reports without developer information.

In this paper, we propose a semi-supervised text classification approach to improve the classification accuracy of bug triage. This semi-supervised approach enhances a NB classifier by applying expectation-maximization (EM) based on the combination of unlabeled and labeled bug reports. First, this semi-supervised approach trains a classifier with labeled bug reports. Then, the approach iteratively labels the unlabeled bug reports and trains a new classifier with labels of all the bug reports. To adjust bug triage, we update such a semi-supervised approach with a weighted recommendation list (WRL) to augment the effectiveness of unlabeled bug reports. This WRL is employed to probabilistically label an unlabeled bug report with multiple relevant developers instead of a single relevant developer. The experimental results on Eclipse indicate that the semi-supervised approach increases the classification accuracy by up to 6%, compared to the original accuracy of 11% to 43% using the supervised NB classifier.

This paper makes the following main contributions:

*1) A semi-supervised text classification approach for bug triage:* We add the unlabeled bug reports to the existing

Our work is partially supported by the Natural Science Foundation of China under Grant No. 60805024, 60903049, the National Research Foundation for the Doctoral Program of Higher Education of China under Grant No. 20070141020, and CAS Innovation Program under Grant No. ISCAS2009-DR01.

supervised approach to avoid the deficiency of labeled bug reports.

*2) A weighted recommendation list for the semi-supervised approach:* We provide a weighted recommendation list for augmenting the semi-supervised approach using probabilistic labels of unlabeled bug reports. Based on this weighted recommendation list, we improve the classification accuracy for the semi-supervised approach.

The remainder of this paper is organized as follows. Section II shows the previous related work of bug triage. Section III presents the semi-supervised bug triage approach and its augmentation of a weighted recommendation list. Section IV shows the experiments on bug reports of Eclipse. In Section V, we discuss the potential problems in the semi-supervised bug triage. Section VI concludes this paper and presents the future work.

## II. RELATED WORK

As to our knowledge, there is no semi-supervised bug triage approach in the literature. All the existing approaches on bug triage and its relevant problems are based on supervised or unsupervised learning. Čubranić & Murphy propose the first work on automatic bug triage [7]. They innovatively apply a supervised learning approach (NB classifier) using text classification to predict relevant developers. They also report the basic steps in the preprocess and provide a set of heuristics for labeling bug reports [2]. Anvik, et al. extend the above machine learning approach [3]. They call their bug triage approach as a semi-automatic approach since they firstly employ a recommendation list to provide the candidate developers for human triagers. Compared with the text classification approaches, Matter, et al. investigate the expertise model of developers on bug triage [12]. Jeong, et al. propose a bug tossing graph approach based on Markov chains from the knowledge of reassigning [11].

The most relevant work of bug triage is detecting duplicate bug reports. In a bug repository, some bug reports are marked as duplicates since such bug reports are just similar as some other handled ones. Runeson, et al. [15] and Wang, et al. [16] remove duplicate bug reports based on supervised natural language processing approaches. Jalbert & Weimer tackle the problem of duplicate bug reports by clustering, an unsupervised learning approach [10]. In contrast to removing the duplicate bug reports, Bettenburg, et al. merge duplicate ones by adding extra information to diagnose actual problems in bug repositories [5].

Before the researches on bug triage, Podgurski, et al. propose a clustering approach to gather the bug reports with the similar errors [14], which can be viewed as the first learning approach for bug reports. They focus on the stack traces in bug reports and analyze the causes of program failures by applying the unsupervised learning approach.

## III. SEMI-SUPERVISED BUG TRIAGE

A reasonable solution for the deficiency of labeled bug reports is to use the unlabeled ones by semi-supervised learning approaches. In semi-supervised classification, we utilize the knowledge from the unlabeled bug reports to assist the existing supervised classifier.

### A. Semi-supervised framework of bug triage

In this paper, we address bug triage by a semi-supervised text classification approach with EM according to the classic text classification approach in [13]. EM is an iterative method, which is used for finding maximum likelihood estimates of parameters in probabilistic models. In our semi-supervised bug triage approach, the classifier with EM fills the "missing values" (labels) of unlabeled bug reports and then trains a new classifier with all the labels of bug reports.

For the application of automatic bug triage, a bug triage approach tests an incoming unlabeled bug report by a trained classifier. To evaluate the effect of a classifier, the data set is divided into two sets: training set for building the classifier and test set for measuring the classification accuracy. In semi-supervised approaches, such a training set is further divided into two subsets: labeled subset with labeled bug reports and unlabeled subset with unlabeled ones.

Algorithm 1 presents the framework of training this semi-supervised approach with EM. There are two basic phases in the framework. One phase is to train a classifier with labeled bug reports; the other phase is EM with both labeled and unlabeled bug reports. The phase of EM iterates two kernel steps. In expectation-step (E-step), the approach evaluates and labels the bug reports in unlabeled subset; in maximization-step (M-step), the approach rebuilds a new classifier with the labels of all the bug reports. The iterations of building classifiers repeat until the performance of classifiers does not improve.

| **Algorithm 1.** Framework of training a semi-supervised classifier with EM |
|---|
| **Input**: labeled subset $R_l$ and unlabeled subset $R_u$ of bug reports, set of developers $D$ <br> **Output**: classifier $\theta$ for semi-supervised bug triage |
| Build a basic classifier $\theta$ supervisedly from bug reports in $R_l$. <br> Loop while classifier $\theta$ improves <br>     E-step. Use classifier $\theta$ to evaluate each bug reports in $R_u$. Label <br>         bug reports in $R_u$. <br>     M-step. Rebulid classifier $\theta$ with bug reports in $R_l$ and $R_u$. |

Most of the supervised approaches can be employed as the basic classifier in the first phase. In this paper, we apply NB classifier due to the following reasons. First, NB is a probability classification framework which can perform well on the text form of bug reports; second, a common method for providing the recommendation is to sort them with probability; third, it is easy to extend EM with a probability weight based on NB.

To enhance the supervised classifier for bug triage, a recommendation list can be employed to enlarge the set of relevant developers. According to a recommendation list of size $n$, the top-$n$ developers can be recommended as the relevant ones instead of only one best developer. This technology is a common strategy on bug triage [1][3][11]. In this paper, we incorporate a WRL into EM to add the weights for multiple relevant developers while training a classifier. The mechanism of these weights provides probabilistic labels for unlabeled bug reports.

*B. NB classifier*

NB classifier is the first approach when treating bug triage as text classification [7]. For further discussion on the semi-supervised approach, we briefly restate the classification framework of NB on bug triage as follows.

Given a set of bug reports $R = \{r_1, r_2, ..., r_{|R|}\}$ and a set of developers $D = \{d_1, d_2, ..., d_{|D|}\}$, the task of bug triage is to assign a relevant developer for an incoming bug report $r$. For a given bug report $r_i$, a classifier $\Phi$ (parameterized on $\theta$) provides a developer $d_j$ which can maximize $P(d_j | r_i; \theta)$ for all $d_j \in D$. Thus, the task of building a NB classifier is to calculate the posterior probability $P(d_j | r_i; \theta)$ and to choose the developer with maximum $P(d_j | r_i; \theta)$. With an application of Bayes' theorem, the posterior probability for a given report $r_i$ is

$$P(d_j | r_i; \theta) = \frac{P(d_j | \theta) P(r_i | d_j; \theta)}{P(r_i | \theta)} \propto P(d_j | \theta) P(r_i | d_j; \theta) \quad (1)$$

The prior probability can be estimated from the training set,

$$P(d_j | \theta) = \frac{\sum_{i=1}^{|R|} P(d_j | r_i)}{|R|} \quad (2)$$

where $P(d_j | r_i) = \{1, 0\}$ from the training set, i.e., if the label of $r_i$ is $d_j$, $P(d_j | r_i) = 1$; else $P(d_j | r_i) = 0$. The list of words is $W = \{w_1, w_2, ..., w_{|W|}\}$ for all the text of bug reports. A NB classifier simplifies the calculation of $P(r_i | d_j; \theta)$ under the assumption that the words are independently and identically distributed (i.i.d.). Thus, the likelihood probability can be estimated as,

$$P(r_i | d_j; \theta) = \prod_{k=1}^{|r_i|} P(w_k | d_j; \theta)^{N_{ki}} \quad (3)$$

where $|r_i|$ is the number of the words in the bug report $r_i$, $N_{ki}$ is the occurrences of word $w_k$ in $r_i$, and

$$P(w_k | d_j; \theta) = \frac{\sum_{i=1}^{|R|} N_{ki} P(d_j | r_i)}{\sum_{m=1}^{|W|} \sum_{i=1}^{|R|} N_{mi} P(d_j | r_i)} \quad (4)$$

To train a classifier, NB estimates (2) and (4) with the training set; to predict a relevant developer for an incoming bug report $r$, all the $P(d_j | r; \theta)$ are calculated with (1) and (3). In practice, a Laplace smoothing is applied to (2) and (4) to avoid zero probabilities. Due to the limitation of paper length, we do not present the formulae after smoothing.

*C. EM in semi-supervised bug triage*

The semi-supervised approach with EM depends on the assumption that the data are generated by a mixture model, and there is a correspondence between mixture components and classes [13]. In semi-supervised triage with EM, some steps of the NB classifier are modified to adapt the unlabeled bug reports.

In E-step, (1) is still used to give the probabilities of labels of bug reports; meanwhile in M-step, the (2) and (4) can be modified to be

$$P(d_j | \theta) = \frac{\sum_{i=1}^{|R|} \Lambda(i) P(d_j | r_i)}{|R_l| + \lambda |R_u|} \quad (5)$$

$$P(w_k | d_j; \theta) = \frac{\sum_{i=1}^{|R|} \Lambda(i) N_{ki} P(d_j | r_i)}{\sum_{m=1}^{|W|} \sum_{i=1}^{|R|} \Lambda(i) N_{mi} P(d_j | r_i)} \quad (6)$$

where $R_l$ and $R_u$ are the labeled subset and the unlabeled subset, respectively. The weight factor of labels is

$$\Lambda(i) = \begin{cases} 1 & \text{if } r_i \in R_l \\ \lambda & \text{if } r_i \in R_u \end{cases} \quad (7)$$

where $\lambda$ is a constant value and $0 \leq \lambda \leq 1$. Obviously, if $\lambda = 0$, the classifier degenerates into a NB classifier. The classifier is a basic EM for $\lambda = 1$, which treats the weights of bug reports in unlabeled subset as those in labeled subset. The classifier changes into a weighted EM for $0 < \lambda < 1$, which treats the bug reports in unlabeled subset with fewer weights than those in labeled subset. The mechanism of weight factor augments the basic EM with the distinction of the labeled and unlabeled bug reports [13]. In practice, the value of $\lambda$ can be estimated by cross-validation method for parameter selection [18]. When the classifier parameters do not improve any more, EM terminates its iterations.

*D. Weighted recommendation list*

Similar as NB classifier with a recommendation list, we design a WRL to guide the M-step of EM. The usage of a recommendation list is to provide a developer list for the decision by a triager; instead, WRL is employed to provide weights for promoting the iterations of EM. With this WRL, we add the weights of multiple developers without maximum posterior probability for bug reports in unlabeled subset.

Algorithm 2 presents the process of training the semi-supervised classifier with a WRL. To implement this extension of EM, E-step recommends the developers with first top-*n* posterior probability for the bug reports in unlabeled subset. In other words, the top-*n* developers are probabilistically labeled for each bug report in unlabeled subset in E-step. The sum of probabilities of developers in the list is one for a bug report. Then in M-step, $P(d_j | r_i)$ is extended as the function $\Gamma(d_j, r_i, q_{ji})$,

$$\Gamma(d_j, r_i, q_{ji}) = \begin{cases} \dfrac{2^{n-q_{ji}}}{\sum_{d_j \in L_i} 2^{n-q_{ji}}} = \dfrac{2^{n-q_{ji}}}{2^n - 1} & \text{if } d_j \in L_i \\ 0 & \text{if } d_j \notin L_i \end{cases} \quad (8)$$

where $L_i$ is a developer list with top-*n* posterior probability of $r_i$ and $q_{ji}$ is the ranking position of the developer $d_j$ for $r_i$ in $L_i$ with $1 \leq q_{ji} \leq n$ (this ranking is based on the posterior probability). For $r_i \in R_l$, the size of recommendation list can be denoted as $n = 1$, i.e., the label of $r_i$ in the labeled subset is still marked as its original relevant developer. Thus, $\Gamma(d_j, r_i, q_{ji}) = 1$, if and only if $d_j$ is the relevant developer of $r_i$ in the labeled subset. With the extension of (8), (5) and (6) can be formed as

$$P(d_j | \theta) = \frac{\sum_{i=1}^{|R|} \Lambda(i) \Gamma(d_j, r_i, q_{ji})}{|R_l| + \lambda |R_u|} \quad (9)$$

$$P(w_k | d_j; \theta) = \frac{\sum_{i=1}^{|R|} \Lambda(i) N_{ki} \Gamma(d_j, r_i, q_{ji})}{\sum_{m=1}^{|W|} \sum_{i=1}^{|R|} \Lambda(i) N_{mi} \Gamma(d_j, r_i, q_{ji})} \quad (10)$$

Thus, in every M-step, the weights of multiple developers are calculated for rebuilding the classifier. In a developer list for an unlabeled bug report, the developer with larger posterior probability can add larger weights for training the classifier. This extension of M-step provides the feasibility for making EM adapt for the semi-supervised bug triage.

---

**Algorithm 2.** Training NB classifier with EM and a WRL

**Input**: labeled subset $R_l$ and unlabeled subset $R_u$,
 set of developers $D$, size $n$ of recommendation list

**Output**: classifier $\theta$ for semi-supervised bug triage

Select $\lambda$ for weight factor of $R_u$ by cross-validation.

Build the NB classifier $\theta$ from the bug reports in $R_l$, while calculating $P(d_j | \theta)$ with (2) and $P(w_k | d_j; \theta)$ with (4).

Loop while classifier $\theta$ does not improve.

 E-step. Use the classifier $\theta$ to estimate the posterior probability of bug reports in $R_u$, $P(d_j | r_i; \theta)$ with (1). Record each list $L_i$ of $r_i$, and $q_{ji}$ for developer $d_j$ in $L_i$.

 M-step. Rebulid classifier $\theta$, with bug reports in $R_l$ and $R_u$, by calculating $P(d_j | \theta)$ with (9) and $P(w_k | d_j; \theta)$ with (10).

---

## IV. EXPERIMENTS

### A. Data preparation

We evaluate the semi-supervised bug triage approach with Bugzilla on Eclipse in the experiments. We take the bug reports with the id from 150001 to 170000 as the data set (The XML form of bug reports can be found in MSR Mining Challenge 2008 [17]). To apply the algorithm described in Section III, it is necessary to preprocess the bug reports as numerical values.

To label the relevant developer for bug reports, a direct method is to identify the developer in assigned-to field. The assigned-to field is a part of a bug report for marking the first relevant developer to fix the bug. This field is filled by triagers when a user of the bug repository submits a bug. However, the developer, who really solves the bug, is not always in agreement with the value of assigned-to field. To solve the problem of this confused field, we follow some previous works [3][7] to label the bug reports with a set of heuristics [2]. Moreover, during the step of labeling bug reports, we remove the bug reports without resolved status, verified status, fixed resolution or duplicate resolution. The status or resolution of a bug report indicates the current status in the life cycle. The removing is not necessary for any bug triage approach (including ours), but we tend to keep the really resolved bug reports. We have 10747 bug reports left after removing. The capability of data set is close to some classic literature [3][7][12].

To extract the numerical values for the bug reports, we select the text of the short description and the first long description to describe a bug report. The reason for this selection is that the triager will face these to describe the details of an incoming bug report [12]. The text form of both short and long description can be tokenized as a list of words and converted into a vector based on words. Thus, every bug report employs such a word vector for recording the counts of words.

Before converting the bug reports, we remove the words in the stoplists and the non-alphabetic tokens to reduce the word vector space [9]. The stoplists store the words with high frequency, which can express little meanings in text. And the non-alphabetic tokens always stand for the specialized words appeared in few bug reports. In addition, we do not use the stemming technology to identify the words with various grammatical suffixes. Many approaches on bug triage employ no stemming since it is not effective in distinguishing bug reports [3][5][7].

After the above steps, numerous words are still left in the word vectors. We remove the words with low frequency to reduce the vector space according to [7]. The words with low frequency can only influence few bug reports and do not provide sufficient information for training a classifier. Similarly, we remove the developers within low frequency to avoid the retired developers according to [3][5][7]. We generated three data sets (in Table I) for comparison after removing the developers who fix less than 10, 30, and 50 bug reports, respectively.

### B. Experimental results

We implement all the approaches with Java (JDK 1.6) in our experiments.

To train and test the classifier effectively, we make the developers in training set, labeled subset and unlabeled subset follow the same probability distribution. For every experiment, the first 5% bug reports of each developer are selected as labeled subset in training set; the following 20% are test set; and the other 75% are unlabeled subset of training set. The unlabeled subset is only used when training a semi-supervised classifier.

In the experimental results, we report the performance of the approaches with the classification accuracy. For a bug triage with a recommendation list, we calculate the accuracy as

$$accuracy_n = \frac{\# \text{ of correct relevant developers}}{\# \text{ of bug reports in test set}}, \quad n \geq 1 \quad (11)$$

where $n$ is the size of the recommendation list. The accuracy is a standard method to measure the performance of bug triage in [5][7][11]. Thus, we do not use other information retrieval metric methods, e.g., precision and recall rates. In our experiments, we abandon the classic evaluation method, $k$-fold cross-validation according to the reasons in [3]. In addition, we select the constant parameter $\lambda$ for the weight factor by cross-validation, which is a standard method of parameter selection in machine learning [18].

We show the classification accuracy while varying the size of the recommendation list on the data sets with different scales in Table I. To simplify the following expression, NBEM is short for NB classifier with EM and NBEM+WRL is short for NBEM with a WRL. We set the size of WRL as the maximum size of recommendation list. From Table I, both NBEM and NBEM+WRL obtain better classification accuracy than NB. For the recommendation list with size 3, this accuracy improves 2% to 5%; for the list with size 5, this accuracy improves 3% to 6%. Considered the original accuracy of NB from 11% to 43%, this improvement is valuable for automatic bug triage. NBEM+WRL also obtains better accuracy than NBEM, from 1% to 3%. It is necessary to note that WRL may hurt performance when the size of the recommendation list is small (e.g., size 2 in the third data set). The reason is that WRL covers the ignored developers in EM, but adds too many weights to the relevant developers in a small recommendation list.

In our semi-supervised approach, EM relies on a constant parameter, which is for the weight factor of unlabeled bug reports. A common method to select such a kind of parameter is cross-validation. It is noted that cross-validation cannot provide the best weight factor for the optimal classification accuracy. In Figure 1, we present the classification accuracy on the third data set in Table I while varying the constant parameter $\lambda$ for the weight factor of the recommendation lists with different sizes. To obtain high classification accuracy, NBEM (Figure 1a) tends to choose the small values near 0.1 for $\lambda$, but NBEM+WRL (Figure 1b) tends to the large values near 1.0. In addition, for a recommendation list with a certain size, the classification accuracy can change up to 3%.

TABLE I. CLASSIFICATION ACCURACY WHILE VARYING THE SIZE OF THE RECOMMENDATION LIST ON THREE DATA SETS WITH DIFFERENT SCALES

| Data set | List size | Accuracy (%) | | |
|---|---|---|---|---|
| | | NB | NBEM | NBEM+WRL |
| 9324 bug reports and 238 developers | 1 | 11.26 | 11.82 | 11.82 |
| | 2 | 17.12 | 17.91 | 18.69 |
| | 3 | 20.78 | 22.07 | 23.48 |
| | 4 | 23.48 | 25.62 | 26.63 |
| | 5 | 26.52 | 28.10 | 30.12 |
| 6965 bug reports and 110 developers | 1 | 15.47 | 16.28 | 16.28 |
| | 2 | 22.65 | 23.32 | 24.50 |
| | 3 | 27.61 | 28.42 | 29.68 |
| | 4 | 31.46 | 32.57 | 33.38 |
| | 5 | 35.09 | 35.97 | 37.75 |
| 5050 bug reports and 60 developers | 1 | 19.92 | 21.04 | 21.04 |
| | 2 | 29.37 | 33.54 | 32.83 |
| | 3 | 34.96 | 39.02 | 39.74 |
| | 4 | 38.52 | 43.50 | 44.92 |
| | 5 | 43.29 | 47.36 | 48.07 |

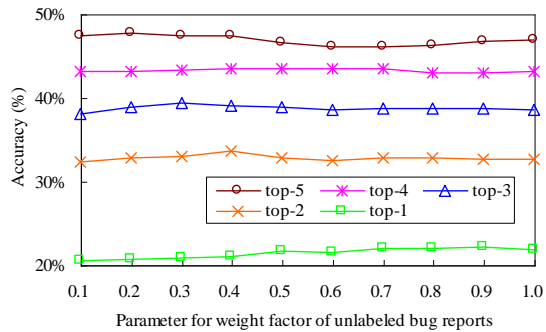

(a) Varying the parameter for weight factor of NBEM

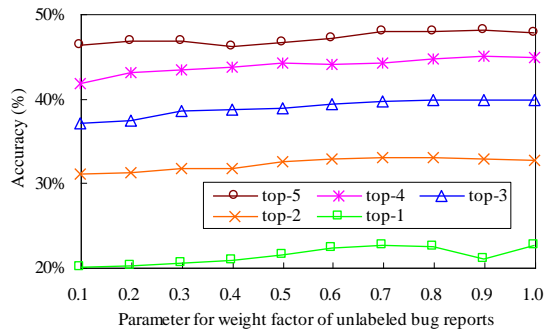

(b) Varying the parameter for weight factor of NBEM+WRL

Figure 1. Classification accuracy while varying the parameter for the weight factor

## V. DISCUSSION

We discuss three topics about our semi-supervised bug triage approach in this section.

The experimental results show that the classification accuracy can be improved by up to 6%. Actually, although the original accuracy of NB is only 11% to 43%, this improvement is not satisfactory for bug triage. The experimental results are insufficient for the real-world applications. As to our knowledge, three reasons are presented as follows: the quality of bug reports is not good enough; a relevant developer for a bug report is hard to label correctly even for human triager; and the mixture model assumption of EM (in Part C of Section III) is not always satisfied in real-world data. In the literature, Čubranić & Murphy mention the basic semi-supervised approach using EM in the discussion of [7], but Anvik, et al. abandon this approach because the results are worse than NB classifier in the discussion of [3]. The reason is that the EM algorithm may hurt classification accuracy when the unlabeled data are in small scale [13]. An extension of the basic EM is a weight factor or a weighted recommendation list, which can partly reduce the dependency of the above mixture model assumption of EM. Moreover, some other extensions may be used to improve the classification accuracy for the semi-supervised approach.

Many bug triage approaches draw an analogy between bug triage and text classification by processing the bug reports based on the existing text process approaches [3][7][15][16]. It is straightforward to take advantage of the text classification approaches in bug triage since most of the content of bug reports consist of free text. However, there are two significant differences between bug triage and text classification. First, the scale of data sets on bug triage is too small in comparison to that on text classification; second, bug triage contains more specialized vocabularies than the common text classification. Thus, not all the text classification approaches can be helpful for the bug triage.

As the basic step of bug triage, we label bug reports under the guideline of the heuristics in [2]. Up till now, these heuristics are the most effective method for labeling. These heuristics can be viewed as a decision tree for automatic labeling bug reports according to the knowledge of the life cycle for bug reports. But these heuristics are hard to implement by programs in traditional bug repositories. A direct solution of this problem is to add a new field to the bug repository in the future to mark a developer who really handles the bug.

## VI. Conclusion and Future Work

Bug triage is a significant step in software development and maintenance. In this paper, we propose a semi-supervised bug triage approach based on NB classifier with EM. This approach improves the classification accuracy with both the labeled and unlabeled bug reports. A WRL is employed to augment the performance via adding the weights of multiple developers when training a classifier. The experimental results demonstrate that our semi-supervised approach improves the classification accuracy of bug triage by up to 6%. During the discussion, we concentrate on three uncertain problems of this approach.

Our future work consists of the following three parts:

*1) Enhancing the classification accuracy via a many-to-one correspondence:* We plan to extend our semi-supervised bug triage approach with EM via the correspondence of many mixture components with one class (many-to-one) proposed in [13]. This correspondence can be considered as the modification of the mixture model of EM for the real-world data. We also plan to combine this many-to-one correspondence with our weighted recommendation list. This combination may improve the accuracy of our semi-supervised bug approach.

*2) Co-training for bug triage:* Apart from naive Bayes with EM, co-training also performs well in semi-supervised learning. Co-training, requiring data with two views, maintains disjoint feature spaces with multiple classifiers on both labeled and unlabeled data. We expect co-training can be applicable for bug triage to avoid the lack of labeled bug reports.

*3) Integration automatic bug triage with a bug repository:* To date, there is no bug triage plug-in component combining with the bug repository in real-world applications. We plan to implement a plug-in component for bug repositories to evaluate automatic bug triage and collect extra information for further researches on bug repositories.

## Acknowledgment

Many thanks to Dr. Thomas Zimmermann with Microsoft Research for sharing bug reports of Eclipse in MSR Mining Challenge 2008. We thank Dr. John Anvik with Department of Computer Science, University of Victoria for sharing the heuristics in labeling bug reports.